\documentclass[aps,prl,twocolumn,superscriptaddress,showpacs,preprintnumbers]{revtex4-1}
\usepackage{natbib}
\usepackage[english]{babel}
\usepackage[dvips]{graphics}
\usepackage{graphicx,epsfig}
\usepackage{amsmath}
\usepackage{color}
\usepackage{multirow}
\usepackage[normalem]{ulem}
\usepackage{booktabs}
\usepackage{physics}
\usepackage{dsfont}
\usepackage{amssymb}
\begin{document}
\title{Optimal quantum reservoir computing for the NISQ era}
%
\author{L. Domingo}
\email[E--mail address: ]{laia.domingo@icmat.es}
\affiliation{Instituto de Ciencias Matemáticas (ICMAT); Campus de Cantoblanco; 
Nicolás Cabrera, 13-15; 28049 Madrid, Spain}
\affiliation{Departamento de Química; Universidad Autónoma de Madrid;
Cantoblanco - 28049 Madrid, Spain}
\affiliation{Grupo de Sistemas Complejos; Universidad Politécnica de Madrid; 28035 Madrid, Spain}

\author{G. Carlo}
\email[E--mail address: ]{carlo@tandar.cnea.gov.ar}
\affiliation{Comisi\'on Nacional de Energ\'ia At\'omica, CONICET, Departamento de F\'isica,
Av.\ del Libertador 8250, 1429 Buenos Aires, Argentina}
\author{F. Borondo}
\email[E--mail address: ]{f.borondo@uam.es}
\affiliation{Instituto de Ciencias Matemáticas (ICMAT); Campus de Cantoblanco;
Nicolás Cabrera, 13-15; 28049 Madrid, Spain}
\affiliation{Departamento de Química; Universidad Autónoma de Madrid;
Cantoblanco - 28049 Madrid, Spain}

\date{\today}
\begin{abstract}
Universal fault-tolerant quantum computers 
require millions of qubits with low error rates.
Since this technology is years ahead,
noisy intermediate-scale quantum (NISQ) computation 
is receiving tremendous interest. 
In this setup, quantum reservoir computing is a relevant machine learning algorithm. 
Its simplicity of training and implementation allows to perform challenging 
computations on today available machines. 
In this Letter, we provide a criterion
to select optimal quantum reservoirs, requiring
few and simple gates. 
Our findings demonstrate that they render better results than 
other commonly used models with significantly less gates, and also provide insight on the theoretical gap between quantum reservoir computing and the theory of quantum states complexity.
\end{abstract}

\maketitle

\textit{Introduction.--} 
Pursuing the early idea of Manin \cite{Manin80} and Feynman \cite{Feynman82} of 
constructing quantum computers, that can solve numerical problems exponentially faster 
than the classical ones, some years ago Google and NASA claimed \cite{google19}, 
not without controversy \cite{IBM19}, having achieved this quantum supremacy.
Although a universal fault-tolerant quantum computer \cite{Barends14}
or efficient real-time error correction \cite{Ryan-Anderson21}
could potentially solve hard benchmark challenging problems, 
such as integer factorization (Shor algorithm \cite{Shor}) or 
unstructured search (Grover's search \cite{grover}), 
and open new perspectives in the applications in many fields \cite{Georgescu14}, 
such as quantum chemistry \cite{QChemistry} or material science \cite{MatSci},
such devices are still decades away from being realised. 
Interestingly enough, this fragility has not decreased the interest in quantum computing, 
but triggered a tremendous activity in the alternative 
called \emph{noisy intermediate-scale quantum} 
(NISQ) era \cite{Preskill18}, 
where quantum algorithms are developed to reach quantum advantage using the 
(small) quantum computers available today \cite{NISQ}. 

A highly relevant NISQ algorithm is \textit{quantum reservoir computing (QRC)} 
\cite{Fujii2021,QRC2} because of its suitability for implementation on NISQ devices. 
QRC has demonstrated to excel not only in classical but also quantum machine learning (QML) tasks. 
It exploits the quantum properties of physical systems and provides an easy training strategy,
achieving excellent results \cite{reviewQRC}. 
In gate-based quantum computation, QRC consists of a random quantum circuit 
applied to an initial state, which encodes the input data,
and the goal is to extract valuable information from the input state, 
so that the measurements of simple local operators are useful features to predict the output.  
These features are then fed to a classical machine learning algorithm, typically a linear model. 
The quantum reservoir (QR) must be a complex quantum circuit, so that the extracted features contain 
enough information for learning the output. Accordingly, the design of the QR is crucial 
for the performance of the model, 
so that selecting optimal QRs is of vital importance. In this respect, the \textit{majorization principle} \cite{majorization_original} has proven to be 
a superb indicator of complexity of random quantum circuits \cite{majorization}. 
Compared to other complexity criteria, such as the entanglement spectrum, evaluating the majorization
in a quantum circuit requires significantly less operations. 
This makes of the majorization principle a criterion {\em specially} suitable for the NISQ era, 
where quantum computation has to be performed with limited quantum resources. 
These limited resources include constraints not only in the size but also regarding the architecture of state of the art quantum processors. Our criterion also directly applies to these qualitatively different technological challenges given that it relies on a global statistical measure rather than on circuit specific proofs. The great relevance of introducing this measure of complexity in the QRC realm is not limited to providing with an extremely efficient way of selecting the optimal QR, but it also bridges the theoretical gap between QRC and the theory of quantum states complexity \cite{linear_growth}. 
As a matter of fact, here we introduce a quantitative classification 
of these algorithms with the growth of quantum circuits complexity, that can be interpreted as the unitaries accessible dimension.

In this Letter, we use the majorization criterion to design the optimal QR in terms of performance for QML. 
The resulting quantum circuits are easily realised in NISQ \cite{NISQ} computers, 
and present a significant advantage over the commonly used Ising model. 
The performance of QRC is assessed using different families of quantum circuits, 
which have different complexity according to the majorization principle. 
Also, we study the number of gates needed for each family to obtain optimal performance. 
In NISQ devices, this number should be as small as possible to minimize error propagation 
due to large error rates and short coherence times \cite{NISQ}. 
We prove that the optimal quantum circuits provided in this work require significantly less
quantum gates that the Ising model, which has been widely used as a QR 
\cite{Fujii2021, QRC2, quantumchemQRC, OptQRC}. 
The optimality of the QR is illustrated by solving a quantum chemistry problem. 
In this context, the data used to train the QRC model is already a quantum state. 
Therefore, it is natural to use a QML algorithm to infer the properties of the system. 
Moreover, to represent a quantum system with $d$ degrees of freedom ($d$ qubits), 
one would need a classical vector of size $2^d$. 
In an actual quantum device, only $d$ physical qubits are needed to encode the system state. 
NISQ devices can currently work with around 100 qubits, which classically would require 
using vectors of a size {em $2^{100}$}, which are totally intractable. 
In this case, the machine learning task is to predict the energy of the excited states of a molecule, 
given its ground state, as presented in \cite{quantumchemQRC}. 
Such ground state can be obtained as an output of another NISQ quantum algorithm, 
such as the  variational quantum eigensolver (VQE) \cite{VQE}. 
This task is relevant since computing the excited states of a Hamiltonian is a much difficult task than computing its ground state. 

\begin{figure*}
\includegraphics[width=0.8\textwidth]{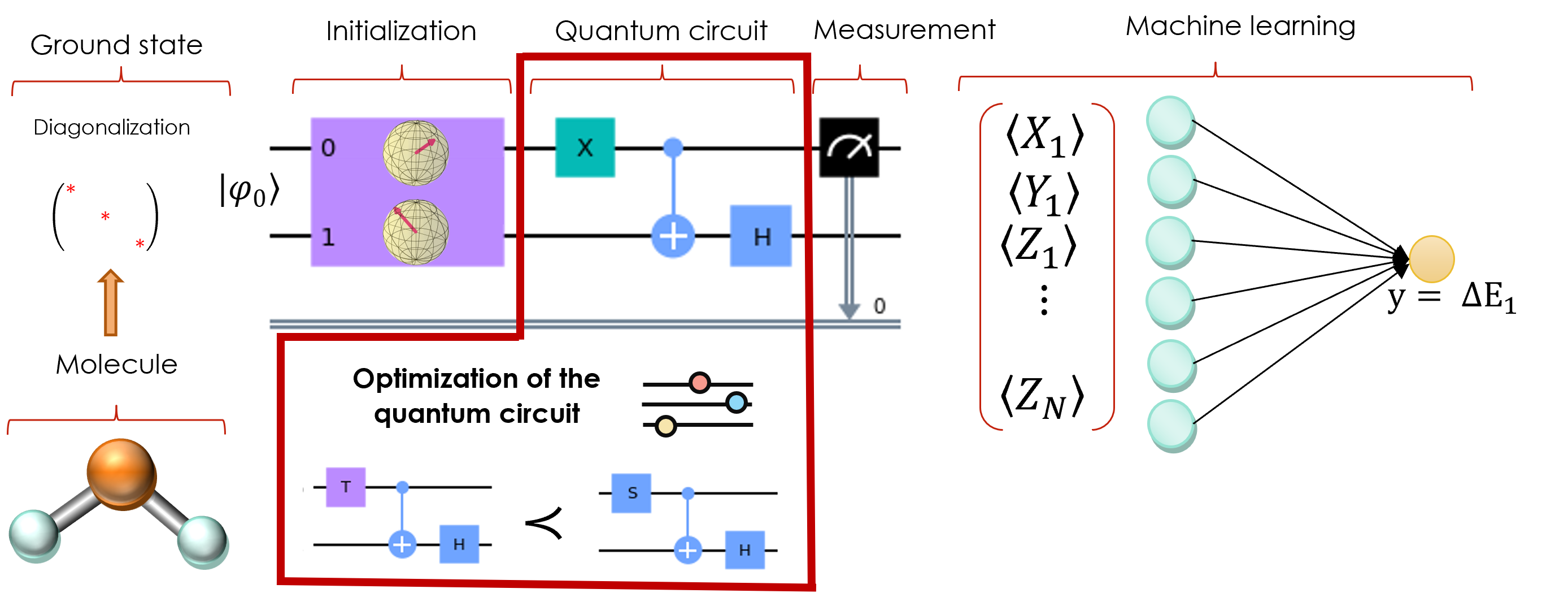}
\caption{Pipeline used to train the quantum reservoir computing model. 
First, the electronic Hamiltonian of the molecule is mapped to the qubit space, 
and its ground state is calculated by direct diagonalization. 
Such ground state is fed to the quantum reservoir, 
which is a random quantum system sampled from one of the seven families studied in this work.
Local Pauli operators are then measured and fed to the classical machine learning algorithm, 
which predicts the excited energy of the molecule. 
The choice of the quantum reservoir is optimised according to the majorization principle
introduced in Ref.~\onlinecite{majorization}. }
\label{fig:1}
\end{figure*}
\textit{Method.--}  The majorization method can be summarizaed as follows.
Let $x, y \in \mathbb{R}^n$. We say that $y$ majorizes $x$ ($y \prec x$) if for all $k<N$
\begin{equation}
    \sum_{i=1}^k x_i^\downarrow < \sum_{i=1}^k y_i^\downarrow, \qquad 
    \sum_{i=1}^N x_i^\downarrow = \sum_{i=1}^N y_i^\downarrow,
\end{equation}
where $x^\downarrow$ indicates that the vector has been arranged in a non-increasing order. 
The partial sums are called cumulants. 
In Ref.~\onlinecite{majorization}, it was shown that the fluctuations of the cumulants 
are a measure of complexity of a quantum circuit. In this work, we consider seven \emph{families} of quantum circuits, which have different complexity according to the majorization principle \cite{majorization}. 
For a given family, the quantum circuit is done by adding random quantum gates from such family. 
We perform 400 simulations for each type of quantum circuit. 
The seven families considerd are the following. 
The first 3 circuits are constructed from a few generators: 
G1 = \{CNOT, H, X\}, G2 = \{CNOT, H, S\}, and G3=\{CNOT,H,T\}, 
where CNOT is the controlled-NOT gate, 
H stands for Hadamard, 
and S and T are $\pi/4$ and $\pi/8$ phase gates, respectively. 
The circuits constructed from G2 generate the Clifford group \cite{G2}, 
and G1 generate a subgroup of Clifford \cite{G1}. 
Therefore, both G1 and G2 are non-universal and classically simulatable. 
On the other hand, G3 is universal and thus approximates the full unitary group $U(N)$ 
to arbitrary precision. 
The third family is composed of Matchgates (MG), which are two-qubit gates formed 
from 2 one-qubit gates, A and B, with the same determinant. 
A acts on the subspace spanned by $\ket{00}$ and $\ket{11}$, while B acts on the subspace 
spanned by $\ket{01}$ and $\ket{10}$. 
A and B are randomly sampled from the unitary group U(2). 
Matchgates circuits are also universal (except when acting on nearest neighbor lines only) \cite{matchgates1, matchgates2}. 
The last families of gates are diagonal in the computational basis. 
As diagonal gates commute, they can be applied simultaneously. 
We separate the diagonal gates into 3 families: D2, D3 and Dn. 
Here, D2 gates are applied to pairs of qubits, D3 gates are applied to 3 qubits, 
and Dn gates are applied to all the qubits. 
The diagonal D2, D3 and Dn families contain $\binom{n}{2}$, $\binom{n}{3}$ and 1 gates, respectively. 
Diagonal circuits cannot perform universal computation but they are not always classically 
simulatable \cite{diagonals}.

With these 7 families of gates quantum circuits suitable as QRs can be designed. 
Apart from the type of gates used, circuits are also characterized by different \emph{number} of gates. 
For the G1, G2, G3 and MG families, we construct circuits of 20, 50, 100, 150 and 200 gates, 
and assess the influence of the number of gates in the final performance of the model. 
For the G3 family, we also construct circuits with up to 1000 gates, 
and with the matchgates we construct circuits with 5, 10 and 15 gates. 
The diagonal circuits have a fixed number of gates, so we only consider that number of gates for these circuits. Additionally, we compare the studied families with the Ising model. 
In this case, the quantum circuit performs the time evolution of a quantum state under the random transverse-field Ising Hamiltonian
\begin{equation}
    H_{\text{Ising}} = \sum_{i,j=0}^{N-1} J_{ij} Z_iZ_j + \sum_{i}^{N-1} h_{i} X_i,
\end{equation}
where $X_i$ and $Z_j$ are Pauli operators acting on the site $i, j$-th qubit, 
and the coefficients $h_i$ and $J_{ij}$ are sampled from the Gaussian distributions 
$N(1, 0.1)$ and $N(0.75, 0.1)$, respectively. 
All time evolutions will be performed for a lapse of time $T =10$. As an illustration, we choose to study the electronic ground and first excited states of two molecules, LiH and H$_2$O in the configuration ranges:
$R_{\text{LiH}} \in [0.5, 3.5]$ a.u., $R_{\text{OH}} \in [0.5,1.5]$ a.u.,
and $\phi_{\text{HOH}}=104.45^\circ$.
The corresponding electronic Hamiltonian and wavefunctions are denoted as 
$H(\vec{R}), \psi_0(\vec{R})$ and $\psi_1(\vec{R})$, respectively, 
The first step is to obtain the ground state of $H(\vec{r})$ in the qubit space. 
To do so, we calculate the second quantization Hamiltonian using the standard STO-3G basis 
for the Hartree-fock optimization \cite{2ndquant1, 2ndquant2}. 
In order to use as few qubits as possible, we remove from the basis set the spin orbitals which
are very likely to be either occupied or virtual in all Slater determinants in the wave function.
In particular, the spin orbitals with a natural orbital occupation number close to 0 or 1 
are assumed empty or occupied, respectively \cite{2ndquant1}. 
The second-quantized Hamiltonian is then mapped to the qubit space by using the Jordan-Wigner transformation \cite{JW}. 
More details about this process are given in the Supplemental Material (SM) \cite{SM}. 
Once the qubit Hamiltonian has been calculated, the corresponding ground state 
$\ket{\psi_0}_{\vec{R}}$ is calculated by (numerical) exact diagonalization. Then, we predict, using the QML algorithm, the target function 
\begin{equation}
  \Delta E(\vec{R}) = E_1(\vec{R}) - E_0(\vec{R}),
\end{equation}
where $E_0(\vec{R})$ and $E_1(\vec{R})$ are the ground state and first excited energies,
respectively. 
In this way, we obtain a 8-qubit ground state for LiH, and a 10-qubit state for H$_2$O. 
For the numerical simulation we split the data set 
$\{\ket{\psi_0}_{\vec{R}}, \Delta E(\vec{R})\}_R$ in  training and test sets. 
The latter contains 30\% of the data, $R_{\text{LiH}} \in [1.1, 2.0]$ a.u.\ and
$R_{\text{OH}} \in [1.05, 1.35]$ a.u.\,
and it is chosen so that the reservoir has to extrapolate to unseen data. 

Once we have generated the training and test data, we design the pipeline of the QRC model,
which is schematically shown in Fig.~\ref{fig:1}. 
The input of the QR, which is a quantum circuit with gates sampled randomly from one of the 
7 gate families described above, is $\ket{\psi_0}_{\vec{R}}$. 
For each experiment the random quantum circuit is the same for all values of $R_{\text{OH}}$ or $R_{\text{LiH}}$.  
After applying the random circuit to the initial state, we measure the expectation values 
of the local Pauli operators $\{X_0, Y_0, Z_0, \cdots, X_n, Y_n, Z_n\}$, 
where $X_i, Y_i, Z_i$ are the Pauli operators $X,Y,Z$ applied to the $i$-th qubit. 
Notice that since the Pauli operators $X_1, \cdots, X_n$ 
(similarly for $Y_1, \cdots, Y_n$ and $Z_1, \cdots, Z_n$) commute with one another, 
the associated observables can be simultaneously measured. 
Therefore, the number of experiments needed to obtain all the observed values does not scale 
with the number of qubits. 
The measurements provide a classical vector $X(\vec{R})$ containing the extracted information 
from the ground state 
\begin{equation}
    X(\vec{R}) = \left( \expval{X_0}, \expval{Y_0}, \expval{Z_0}, \cdots,  
        \expval{X_n}, \expval{Y_n}, \expval{Z_n} \right)^T ,
\end{equation}
where $n$ is the number of qubits, and the expectation values are taken over $\ket{\psi_0}_{\vec{R}}$.
This classical vector is then fed to a classical machine learning model. 
In this work we use the ridge regression, a regularized linear model which minimizes 
the mean squared error
\begin{equation}
    \text{MSE}_R = 
    \frac{1}{N_s} \sum_{i=0}^{N_s} \left[ W \cdot X(\vec{r}_i) - y(\vec{r}_i) \right]^2 
    + \alpha ||W||^2
\end{equation}
where $N_s$ is the number of samples in the training set, 
$W$ is the matrix of the linear model, $\alpha$ is the regularization parameter, 
and $||\cdot||$ is the $L^2$ norm. 
Notice that since the linear model has to extrapolate to unseen data 
(unseen values of $\vec{R}$), it is necessary to add regularization to the learning algorithm 
in order to prevent overfitting the training data. 
In this work we choose $\alpha=10^{-7}$, a value that simultaneously prevents overfitting 
and provides accurate predictions.
Although it is true that any other classical machine learning algorithm could have been 
used, instead of this linear model, at this point,
the QR is able to extract valuable information from the quantum state.
Thus, a simple machine learning model, like the one we are using here, 
is plenty enough to predict the excited properties of the system.  

%
\begin{figure}
 \includegraphics[width=1.0\columnwidth]{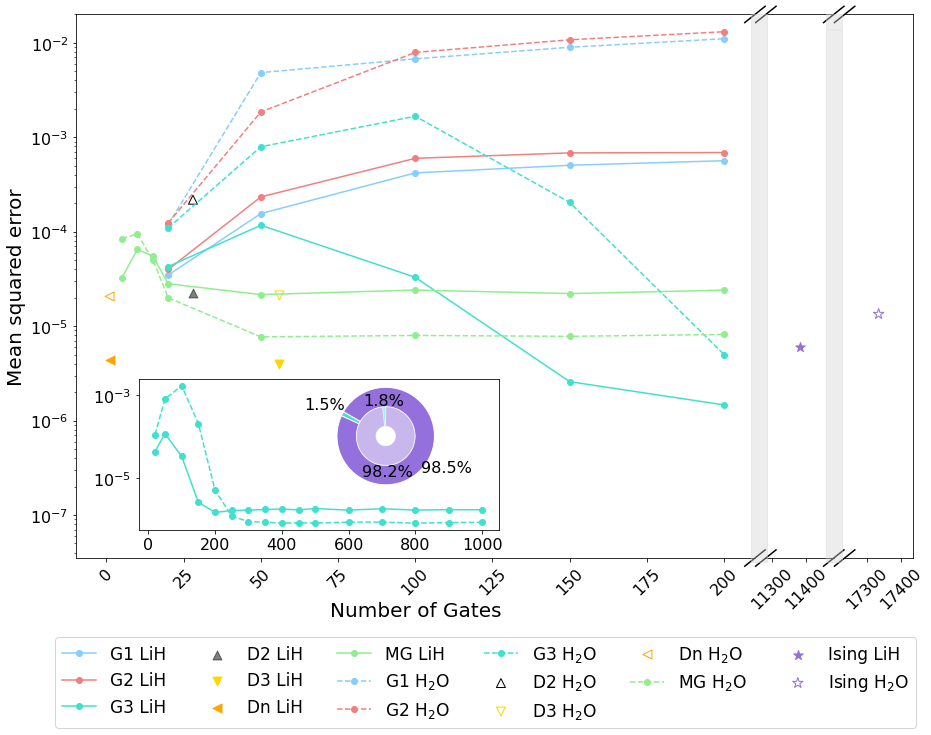}
  \caption{Average mean squared error (MSE) of the seven families of quantum reservoirs 
  as a function of the number of gates of the circuit. 
  Averages are made over 400 simulations. 
  The solid and dashed lines represent the results for the LiH and H$_2$O molecules respectively. 
  The plot also displays, for comparison, the average performance of the Ising model. 
  The inset contains the MSE of the G3 circuits for larger number of gates. 
  The pie charts represents the proportion of gates needed to obtain optimal performance 
  for the G3 family and the Ising model. }
 \label{fig:2}
\end{figure}
\par
\textit{Results.--} Figure~\ref{fig:2} shows the performance of QRC for the 7 families 
of quantum circuits as a function of the number of gates of the circuits. 
Solid lines correspond to the LiH molecule and dashed lines to H$_2$. 
As can be seen, the performance of the different circuits is qualitatively the same 
in both cases. 
In Fig.~3 of Ref.~\onlinecite{majorization}, it is seen how the fluctuations of the 
Lorentz curves differentiate the various families of random circuits,
with the families with lower fluctuationscloser to the Haar measure behavior corresponding to more complex random quantum circuits.
Our results in Fig~\ref{fig:2} agree with the classification in given in 
Ref.~\onlinecite{majorization}. 
Circuits with higher complexity have a better performance in the QML task. 
In particular, the G1 and G2 families are the less complex families according to the 
majorization indicator, and they also give worse results in the QML task. 
On the other hand, the circuits in the G3 family give the lowest error in the predictions, 
which agrees with the G3 family having the highest complexity. 
The matchgates circuits, on the other hand, have a slightly worse performance than the G3 family, 
which also agrees with the majorization criterion. 
Finally, the performances of the D3 and Dn circuits are very similar to each other, 
although the performance of the D2 circuits is significantly worse. 
This difference between the diagonal circuits is also seen when using the majorization indicator.  
Figure~\ref{fig:2} also shows how the performance of the QML task changes with the number 
of gates of the circuit.  
The G1 and G2 circuits give worse results as the number of gates increase. 
On the contrary, the G3 circuits improve its performance with the number of gates, 
and this performance stabilises around 200 gates for LiH, and around 250 for H$_2$O. 
The matchgate circuits also improve its performance with the number of gates, 
but in this case the optimal performance is achieved with only 20 gates (LiH) or 50 gates (H$_2$O),
even though this optimal error is higher than the optimal error of G3. 
Notice that the H$_2$O system is larger (10 qubits) than for the LiH system (8 qubits). 
Therefore, predicting the excited energy, $E_1(\vec{R})$, for the H$_2$O is a harder task, 
and it is expected that the optimal circuit requires more gates. 
The same analysis has been performed to predict the second excited energy, $E_2(\vec{R})$,
for the two molecules under study. 
The results are qualitatively the same and are provided in the SM \cite{SM}. 

\begin{figure}
  \includegraphics[width=0.95\columnwidth]{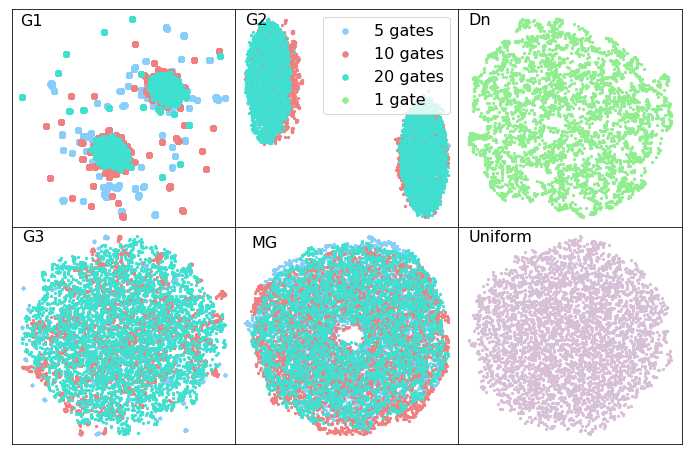}
  \caption{2D representation of the distribution of the families of quantum circuits in the Pauli space, compared to the uniform distribution.}
\label{fig:3}
\end{figure}

To further understand the difference in the performance of the circuits,  
we inspect how each of the random circuits span the space of operators. 
For simplicity and easier visualisation we create a toy model of two qubits and apply 
random circuits from each of the families. 
Each of the circuits is a unitary operator $U$, constructed by the successive application 
of the gates of the circuit. 
This operator can be written as a linear combination of the elements of the Pauli space 
$\{\mathds{1} \otimes \mathds{1}, \mathds{1} \otimes X, \mathds{1} \otimes Y, 
\mathds{1} \otimes Z, \cdots, X\otimes Z, Y \otimes Z, Z \otimes Z \}$. 
For each family of gates, we design 4000 random circuits and see how their unitaries 
fill the Pauli space, compared to the uniform distribution. 
Since the Pauli space in the 2-qubit system is a 16-dimensional space, 
we use a dimensionality reduction technique called UMAP \cite{UMAP2} to visualise the 
distribution in 2D. 
The details of this algorithm are provided in the Supplemental Material \cite{SM}. 
The results are shown in Fig.~\ref{fig:3}.  
We see that the G1 and G2 circuits only fill a subset of the Pauli space. 
Moreover, when the number of gates is small, the filling of the space is more sparse. 
As the number of circuit gates increase, the unitaries concentrate in a dense region of the space. 
This fact justifies that QRs with more gates produce worse results in the QML task 
for these two families. 
On the other hand, the G3 family fills the Pauli space uniformly. 
As the number of gates increase, the distribution of the unitaries resembles 
more the uniform sampling. 
For this reason, the performance of the QR improves with the number of gates, 
until it achieves its optimal value. 
Regarding the matchgate circuits, they behave similarly to the G3 family, 
except that they do not fill a small region of the Pauli space, 
this leading to a slightly worse performance in the QML algorithm. 
Finally, the diagonal circuits (which in the 2-qubit space coincide) also fill the 
whole Pauli space. 
However, there are regions with higher density than others. 
This also illustrates the slightly higher error in the QRC model. 
The SM \cite{SM} contains a short video complementing Fig.~\ref{fig:3}. 
In it, the number of gates of the circuit change in time. 
After analysing the performance of the 7 families of circuits, 
we can compare them with the results of the Ising model, which has been extensively used 
as QR \cite{QRC, QRC2, quantumchemQRC, OptQRC}. 
Figure~\ref{fig:2} shows that the MSE of the Ising model is slightly higher than the MSE 
for the G3 family. 
Apart from its performance, we can also calculate the number of gates needed to implement 
the Ising model in a gate-based quantum computer. 
The time evolution operator $e^{iH_{\text{Ising}}T}$ can be approximated by first order 
Trotter decomposition \cite{Trotter1}. 
This decomposition provides a quantum circuit with gates from the set 
\{H, CNOT, $R_z(\theta_i)\}_i$, where $R_z(\theta_i)$ is the rotation of an angle $\theta_i$ 
around the Z axis. 
Unfortunately, it is impossible to implement rotations with perfect accuracy in the 
current quantum computers. 
Fault-tolerant quantum computers typically perform these rotations using multiple applications 
of gates H and T. 
For this reason, we have decomposed \cite{HT} the Ising time evolution operator to a 
quantum circuit with gates from the set $\{$H, CNOT, T$\}$, which is the G3 family. 
We have run 400 simulations with the different parameters in the Ising Hamiltonian. 
For the LiH molecule, constructing the Ising quantum circuit required on average 
11381 gates, and 17335 for H$_2$O. 
As shown in Fig.~\ref{fig:2}, the optimal performance of the QR for the G3 family is 
achieved with only 200 gates (250 for H$_2$O). 
Therefore, there is no need to use tens of thousands of gates to achieve optimal performance 
in the QR.

\textit{Conclusions.--} In this Letter we provide a criterion to design optimal QRs for QML. 
This criterion is also easy to implement in NISQ devices, 
since it requires using circuits with only few quantum gates. 
We have demonstrated that the optimal circuits obtained with the prescriptions in this work give
better performance than the commonly used Ising model, and they need significantly less quantum gates.
This is of outmost importance for optimal implementations in current NISQ devices, 
since only circuits with few and simple gates can guarantee the required accuracy.
The key point of our work is the use of the majorization principle \cite{majorization} 
as an indicator for the complexity of the quantum circuits. 
The QRs with higher complexity according to the majorization criterion provide better results 
in QML tasks. 
We also give an intuitive explanation of how the studied families of quantum circuits 
extract information from their initial state. 
We see that the optimal family of quantum circuits fills uniformly the Pauli space of operators, 
while the other families do not reproduce all the operators in the Pauli space. 

\textit{Code Availability Statement.--}
The code and data that support the findings of this study are openly available in \href{https://github.com/laiadc/Optimal\_QRC}{https://github.com/laiadc/Optimal\_QRC}.

\textit{Acknowledgments.}
The project that gave rise to these results received the support of a fellowship from "la Caixa" Foundation (ID 100010434). 
The fellowship code is LCF/BQ/DR20/11790028.
This work has also been partially supported by the Spanish Ministry of Science, Innovation and Universities, 
Gobierno de Espa\~na, under Contracts No.\ PGC2018-093854-BI00, ICMAT Severo Ochoa CEX2019-000904-S;
and by the People Programme (Marie Curie Actions) of the European Union's Horizon 2020 Research and Innovation Program 
under Grant No.~734557.

\bibliography{bibliography}

\end{document}